\documentclass[twocolumn,9pt]{article} 

\usepackage[square,numbers,sort&compress,comma]{natbib}

\usepackage{amsmath}
\usepackage{amssymb}
\usepackage{caption}
\usepackage{graphicx}
\usepackage{latexsym}
\usepackage{times}
\usepackage[pagewise]{lineno}
\usepackage{hyperref}
\usepackage{chemformula}
\usepackage{booktabs}
\usepackage{subcaption}

\topmargin - 12pt 
\renewenvironment{abstract}%
              {
               \small
               {\bfseries \abstractname}
               \par
               \vspace{10pt}
              }

\renewcommand\abstractname{Abstract}

\newcommand{\nomenclature}
              [1]
              {
               \bgroup
               \flushleft
               \small\bf
               #1
               \par
               \egroup
              }

\renewcommand{\section}
              [1]
              {
               \bgroup
               \flushleft
               \small\bf
               \refstepcounter{section}
               \arabic{section}. #1
               \par
               \egroup
              }

\renewcommand{\subsection}
              [1]
              {
               \bgroup
               \flushleft
               \small\em
               \refstepcounter{subsection}
               \arabic{section}.
               \arabic{subsection}. #1
               \par
               \egroup
              }

\renewcommand{\subsubsection}
              [1]
              {
               \bgroup
               \flushleft
               \small\em
               \refstepcounter{subsubsection}
               \arabic{section}.
               \arabic{subsection}.
               \arabic{subsubsection}. #1
               \par
               \egroup
              }

  \newcommand{\acknowledgement}
              [1]
              {
               \bgroup
               \flushleft
               \small\bf
               #1
               \par
               \egroup
              }

  \newcommand{\sectionbib}
              [1]
              {
               \bgroup
               \flushleft
               \small\bf
               #1
               \par
               \egroup
              }

\begin{document}



\small
\baselineskip 10pt

\setcounter{page}{1}
\title{\LARGE \bf Effect of fuel stratification length scale on thermodiffusively unstable lean hydrogen flames}

\author{{\large Filippo Fruzza$^{a,*}$, Sofiane Al Kassar$^{b}$, Rachele Lamioni$^{a}$,}\\
        {\large Chiara Galletti$^{a}$, Antonio Attili$^{b}$}\\[10pt]
        {\footnotesize \em $^a$Department of Civil and Industrial Engineering, University of Pisa, 56122 Pisa, Italy}\\[-5pt]
        {\footnotesize \em $^b$School of Engineering, Institute for Multiscale Thermofluids, The University of Edinburgh,}\\[-5pt]
        {\footnotesize \em EH93FD Edinburgh, United Kingdom}\\[-5pt]}

\date{}  

\twocolumn[\begin{@twocolumnfalse}
\maketitle
\rule{\textwidth}{0.5pt}
\vspace{-5pt}

\begin{abstract}
Lean premixed hydrogen flames are highly susceptible to thermodiffusive instabilities, which generate complex cellular structures and significantly enhance propagation speed. In practical combustors, incomplete premixing can introduce spatial variations in local equivalence ratio upstream of the flame. This work investigates how the characteristic length scale of inlet fuel stratification affects the structure and propagation of lean laminar hydrogen flames using direct numerical simulations with detailed chemistry and transport. Controlled sinusoidal perturbations of the fuel mass fraction are imposed at moderate amplitude, leading to local equivalence-ratio variations of approximately $\phi\simeq0.45$--$0.55$ in the flame region. Seven stratification wavelengths, from $\lambda=15.4\,\delta_f$ to $133.3\,\delta_f$, are examined while maintaining a constant density-weighted global equivalence ratio. Additional unity-Lewis-number simulations are used to separate the purely geometrical response to mixture inhomogeneity from the effects of differential transport. The results reveal a strongly scale-dependent response. In unity-Lewis-number flames, stratification organises the front at the imposed wavelength and produces a monotonic increase of flame surface area and propagation speed with increasing wavelength. In flames with differential diffusion, large wavelengths ($\lambda\gtrsim40\,\delta_f$) produce a similar large-scale organisation, with rich channels forming forward bulges and lean channels forming recessed cusps, while thermodiffusive cells remain superimposed on this imposed structure. This increases the overall flame surface and enhances the global propagation speed. At smaller wavelengths ($\lambda\lesssim29\,\delta_f$), instead, the imposed composition gradients interfere with the intrinsic thermodiffusive cellular dynamics, reducing flame surface area and global propagation speed relative to the perfectly premixed reference. These findings identify stratification length scale as a key parameter controlling the interaction between moderate mixture inhomogeneity and thermodiffusive instability in lean hydrogen flames.
\end{abstract}

\vspace{10pt}

{\bf Novelty and significance statement}

\vspace{10pt}

This work presents the first systematic DNS investigation of how inlet fuel stratification affects thermodiffusively unstable lean premixed \ch{H2}/air flames. Controlled sinusoidal mixture inhomogeneities are imposed while maintaining a constant density-weighted global equivalence ratio. \textcolor{black}{The study identifies the stratification wavelength as a scale-selection parameter that can either suppress or reinforce thermodiffusive flame structures. When the imposed wavelength overlaps with the intrinsic cellular size range, transverse composition gradients disrupt the self-organised thermodiffusive cells, reducing flame surface area and global burning rate. For larger wavelengths, the flame behaves similarly to the corresponding unity-Lewis-number stratified flames at large scale, forming alternating rich and lean channels, while thermodiffusive cells remain active at smaller scales. This produces a net increase in flame surface area and propagation speed.} A passive tracer is introduced as a diagnostic tool to track the imposed stratification independently of combustion chemistry and differential diffusion. \textcolor{black}{Tracer-conditioned statistics show that the dominant mechanism is the scale-dependent redistribution of flame surface, rather than a systematic enhancement of local burning rates.}

\vspace{5pt}
\parbox{1.0\textwidth}{\footnotesize {\em Keywords:} Hydrogen flames; Thermodiffusive instability; Fuel stratification; Mixture inhomogeneity; DNS}
\rule{\textwidth}{0.5pt}
*Corresponding author.
\vspace{5pt}
\end{@twocolumnfalse}] 

\section{Introduction\label{sec:intro}}\addvspace{10pt} 

Lean premixed \ch{H2}/air combustion is of growing interest for low-emission gas-turbine applications, but also poses challenges due to hydrogen’s high reactivity and broad flammability limits. \textcolor{black}{Premixed flames are subject to intrinsic instabilities arising from hydrodynamic and thermal-diffusive mechanisms~\cite{matalon2007,kadowaki2005}. In lean \ch{H2}/air mixtures, the low Lewis number of hydrogen enhances thermodiffusive effects, leading to flame-front corrugation, cellular structures, and increased global burning rates~\cite{altantzis2012,frouzakis2015}. Previous numerical studies have shown that the nonlinear dynamics of these flames depend on key parameters such as Lewis number, equivalence ratio, pressure, and domain size~\cite{altantzis2012,frouzakis2015,berger2019,creta2020,howarth2022,berger2022a,berger2022b,howarth2023,berger2023,wen2024a,wen2024b}.}

\textcolor{black}{In practical systems, however, the mixture is rarely perfectly premixed at the burner inlet. Incomplete mixing, due to injector design or finite residence times, introduces spatial variations in equivalence ratio upstream of the flame~\cite{helie1998,jimenez2002,zirwes2021}. While intrinsic instabilities in homogeneous mixtures are well documented, the response of thermodiffusively unstable hydrogen flames to imposed mixture inhomogeneities remains less explored. In particular, the interaction between the stratification length scale and the intrinsic wavelength of thermodiffusive instabilities is still not fully understood~\cite{chu2025}. This interaction is expected to be important, as fixed external length scales have been shown to influence the dynamics of intrinsically unstable flames, for instance through domain-size effects that constrain the development of unstable wavelengths~\cite{creta2020,lapenna2021proci,lapenna2021CTM}.}

The present work investigates how the spatial scale of inlet fuel stratification influences the structure and propagation of lean premixed hydrogen flames. Direct numerical simulations (DNS) with detailed chemistry and transport are performed, with controlled sinusoidal perturbations of the fuel mass fraction imposed at the inlet to systematically vary the stratification wavelength $\lambda$. The density-weighted fuel input is kept constant across all cases, ensuring that the observed differences arise solely from the stratification wavelength rather than from changes in the global equivalence ratio.
\textcolor{black}{Seven stratification wavelengths ranging from $\lambda = 15.4\,\delta_f$ to $\lambda = 133.3\,\delta_f$, where $\delta_f$ is the laminar flame thickness, are examined at moderate modulation amplitudes}. The resulting flame dynamics are analysed in terms of instantaneous structure and global propagation characteristics to identify how fuel stratification interacts with intrinsic thermodiffusive instabilities. The results provide new insight into how the competition between stratification scale and instability wavelength governs the behaviour of lean \ch{H2}/air flames, with implications for the design and operation of partially premixed hydrogen combustors.

\section{Physical models and numerical methods\label{sec:models}}\addvspace{10pt}

The governing equations are the reactive, unsteady Navier-Stokes equations solved under the low-Mach-number approximation, with the mixture following the ideal gas equation of state. A finite-rate multistep chemistry model involving 9 species and 23 reversible reactions~\cite{burke2012comprehensive} is employed for the species and temperature equations, as described in~\cite{attili2016effects}. Transport properties are computed using a mixture-averaged model~\cite{attili2016effects}, and the thermodiffusion (Soret) effect is included via the model of Schlup and Blanquart~\cite{SCHLUP20181}.

The equations are discretised using a semi-implicit finite-difference method~\cite{desjardins_high_2008}, previously validated in various configurations~\cite{berger2019,berger2022a,berger2022b,attili2021}. Second-order finite differences are used for momentum and scalar diffusive terms, while a third-order WENO scheme~\cite{liu1994} is applied to scalar convective terms. The chemical source term is integrated via operator splitting~\cite{strang1968} with the stiff ODE solver \texttt{CVODE}~\cite{hindmarsh2005}.

\section{Configuration and inflow fuel stratification strategy\label{sec:config}}\addvspace{10pt}

\textcolor{black}{The simulations are performed in rectangular domains with two different sizes. The first has dimensions $L_x\times L_y=200\,\delta_f\times200\,\delta_f$ and resolution $N_x\times N_y=2048\times2048$, while the second has dimensions $L_x\times L_y=400\,\delta_f\times300\,\delta_f$ and resolution $N_x\times N_y=4096\times3072$. This choice ensures a uniform grid spacing across all cases, corresponding to approximately $10$ points per laminar flame thickness $\delta_f$, while allowing the domain size to be adapted to the imposed stratification scale. A mesh-independence study up to $\sim15$ points per $\delta_f$ was performed for a representative case and confirmed a limited sensitivity of both global and local statistics to further grid refinement; the corresponding results are reported in the Supplementary Material.} Periodic boundary conditions are imposed in the spanwise direction ($x$), while inflow/outflow conditions are applied in the streamwise direction ($y$). \textcolor{black}{This setup is consistent with similar DNS configurations for thermodiffusively unstable premixed flames~\cite{berger2019,berger2022a,howarth2022,howarth2023,creta2020}.}

To investigate the effect of mixture inhomogeneity on flame dynamics, controlled spatial inhomogeneity is introduced at the inlet through a transverse sinusoidal modulation of the hydrogen mass fraction:
\begin{equation}
    Y_{H_2}(x) = \hat{Y}_{\ch{H2}}\,[\,1 + A \sin(2\pi\,x/\lambda)\,],
\end{equation}
where $\hat{Y}_{\ch{H2}}$ is the mean hydrogen mass fraction, $A$ is the fluctuation amplitude, and $\lambda$ is the wavelength of the perturbation. \textcolor{black}{At the inlet, a uniform normal velocity is imposed across the transverse direction, while only the composition is spatially modulated according to Eq.~(1).} To preserve periodicity at the inlet, $\lambda$ is selected such that an integer number of complete oscillations fits within the transverse domain length, i.e., $\lambda = L_x/n$ with $n \in \mathbb{N}$. The remaining mass fraction is assigned to air while preserving the \ch{O2}/\ch{N2} ratio.

Because mixture density varies with composition at constant pressure $p$ and temperature $T$, the globally injected composition differs from the spatial average. The effective hydrogen mass fraction is therefore defined as the density-weighted mean:
\begin{equation}
Y_{H_2}^{\mathrm{eff}} =
\frac{\int_0^{L_x} \rho\!\left(Y_{H_2}(x)\right)\,Y_{H_2}(x)\,dx}
{\int_0^{L_x} \rho\!\left(Y_{H_2}(x)\right)\,dx},
\end{equation}
where $\rho$ denotes the mixture density, which depends on the local composition $Y_{H_2}(x)$ under the imposed constant pressure and temperature conditions. For the sinusoidal profile, introducing
\begin{equation*}
\begin{gathered}
    \alpha = \frac{1}{M_{\mathrm{air}}} + \left(\frac{1}{M_{H_2}} - \frac{1}{M_{\mathrm{air}}}\right)\hat{Y}_{\ch{H2}},\\
    \beta = \left(\frac{1}{M_{H_2}} - \frac{1}{M_{\mathrm{air}}}\right)\hat{Y}_{\ch{H2}}A,  
\end{gathered} 
\end{equation*}
where $M_{H_2}$ is the molecular weight of hydrogen and $M_{\mathrm{air}}$ is the mean molecular weight of air at the prescribed \ch{O2}/\ch{N2} ratio, ${1/M_{\mathrm{air}}=0.233/M_{\ch{O2}}+0.767/M_{\ch{N2}}}$. This yields the closed-form expression
\begin{equation}
Y_{H_2}^{\mathrm{eff}} = \hat{Y}_{\ch{H2}}\,\Bigl[\,1 + A\,\frac{\sqrt{\alpha^{2}-\beta^{2}} - \alpha}{\beta}\,\Bigr].
\end{equation}
In practice, $\hat{Y}_{\ch{H2}}$ is adjusted to match the target global equivalence ratio $\phi$ at the desired amplitude $A$.

The reference flame is a premixed \ch{H2}/air mixture at $p=20$~atm, unburnt temperature $T_u=700$~K, and equivalence ratio $\phi=0.5$, conditions under which strong thermodiffusive instability occurs. The flame is initialised with multi-wavelength perturbations to accelerate transition to a fully developed state~\cite{alkassar2024}. Once transients decay and fully developed thermodiffusive instabilities are established, the sinusoidal inlet perturbation is activated. The mean inlet velocity is adjusted to balance the flame's propagation speed, keeping the time-averaged flame position stationary in the domain. \textcolor{black}{The effect of the imposed inlet wavelength is investigated by considering seven stratification scales, from $\lambda=15.4\,\delta_f$ to $\lambda=133.3\,\delta_f$, in addition to the perfectly premixed reference. The four smallest-wavelength cases, $\lambda=15.4$, $20$, $28.6\,\delta_f$, and $40\,\delta_f$, are computed in the $200\,\delta_f\times200\,\delta_f$ domain, while the larger-wavelength cases are computed in the $400\,\delta_f\times300\,\delta_f$ domain to fully capture the larger flame structures induced by the imposed stratification. In all stratified cases, moderate modulation amplitudes are considered, with $A=0.2$ for $\lambda=15.4\,\delta_f$ and $\lambda=20\,\delta_f$, and $A=0.1$ for all larger wavelengths, leading to local equivalence-ratio variations of approximately $\phi\simeq0.45$--$0.55$ in the flame region. This choice allows isolating the effect of the stratification length scale, since appreciable changes in flame topology and propagation speed already occur without introducing the stronger nonlinear coupling expected for larger mixture variations. Stratification wavelengths smaller than $\lambda=15.4\,\delta_f$ were not considered, as the associated transverse gradients lead to rapid diffusive smoothing of the imposed inhomogeneity before reaching the flame front (see Section~\ref{sec:dynamics}). The perfectly premixed reference was computed in both domains, with no appreciable differences in the statistically stationary flame dynamics or global consumption speed, consistently with the domain-size independence demonstrated in~\cite{berger2019}.}

\textcolor{black}{To assess the role of differential transport, additional unity-Lewis-number simulations are performed for all stratified wavelengths, using the corresponding domain, grid resolution, and inlet stratification parameters as in the non-unity Lewis number simulations. For these unity-Lewis-number cases, $\delta_f$ and $s_L$ denote the laminar flame thickness and the unstretched laminar flame speed computed from the corresponding one-dimensional $Le=1$ reference flame.} 

Table~\ref{tab:cases} summarises the computational domains, stratification parameters, and corresponding inlet velocities required for flame stabilisation in each case.

\begin{table}[!ht]\footnotesize
\caption{Summary of the stratification parameters and corresponding inlet velocities used to stabilise the flame. $n$ denotes the number of oscillations at the inlet, $\lambda$ the imposed wavelength, $L_x$ the spanwise domain width, $A$ the modulation amplitude, and $V_{\text{in}}/s_L$ the inlet velocity normalised by the unstretched laminar flame speed. MA and UL denote mixture-averaged and unity-Lewis-number transport, respectively.}
\centering
\setlength{\tabcolsep}{3.5pt}
\begin{tabular}{llccccc}
\hline 
Case & Trans. & $n$ & $\lambda/\delta_f$ & $L_x/\delta_f$ & $A$ & $V_{\text{in}}/s_L$ \\
\hline
Ref   & MA & 0  & $\ $   & 200 \& 400 & $\ $ & 2.12\\
MA15  & MA & 13 & 15.4   & 200 & 0.2  & 1.80\\
MA20  & MA & 10 & 20.0   & 200 & 0.2  & 1.82\\
MA29  & MA & 7  & 28.6   & 200 & 0.1  & 1.97\\
MA40  & MA & 5  & 40.0   & 200 & 0.1  & 2.27\\
MA67  & MA & 6  & 66.7   & 400 & 0.1  & 2.47\\
MA100 & MA & 4  & 100.0  & 400 & 0.1  & 2.49\\
MA133 & MA & 3  & 133.3  & 400 & 0.1  & 2.60\\
UL15  & UL & 13 & 15.4   & 200 & 0.2  & 1.08\\
UL20  & UL & 10 & 20.0   & 200 & 0.2  & 1.09\\
UL29  & UL & 7  & 28.6   & 200 & 0.1  & 1.13\\
UL40  & UL & 5  & 40.0   & 200 & 0.1  & 1.16\\
UL67  & UL & 6  & 66.7   & 400 & 0.1  & 1.28\\
UL100 & UL & 4  & 100.0  & 400 & 0.1  & 1.27\\
UL133 & UL & 3  & 133.3  & 400 & 0.1  & 1.28\\
\hline
\end{tabular}
\label{tab:cases}
\end{table}

\section{Results\label{sec:results}}


\subsection{Dynamical response to inlet fuel stratification\label{sec:dynamics}}\addvspace{10pt}

To visualise the mixture inhomogeneity generated by the inlet stratification, the Bilger mixture fraction~\cite{bilger1989} field is expressed in terms of its relative deviation from the inlet reference value, $z = \frac{Z}{Z_0} - 1$, where $Z_0$ denotes the Bilger mixture fraction at the inlet of the perfectly premixed reference case. This normalisation highlights the spatial modulation of the mixture composition while removing the uniform background value. Positive values of $z$ therefore correspond to locally fuel-rich regions relative to the reference mixture, whereas negative values indicate fuel-lean regions.

\begin{figure}[!ht]
    \centering
    \includegraphics[width=0.9\columnwidth]{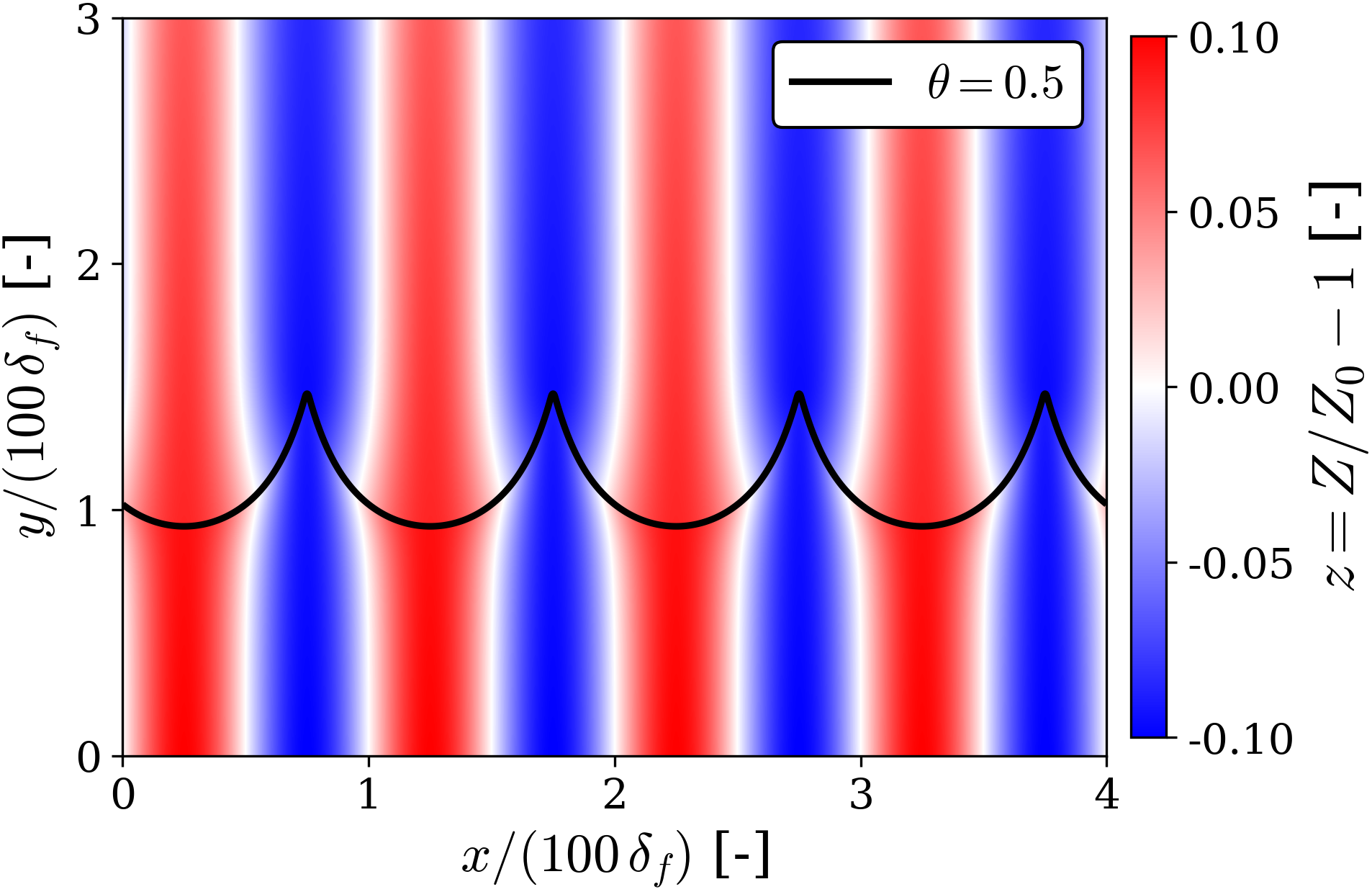}
    \caption{Instantaneous normalised Bilger mixture fraction field, $z=Z/Z_0-1$, for a unity-Lewis-number flame with imposed inlet stratification. The black line denotes the $\theta=0.5$ isocontour. The case shown corresponds to $\lambda=100\,\delta_f$.}
    \label{fig:le1_stratified}
    \vspace{-10pt}
\end{figure}

\begin{figure*}[!t]
    \centering
    \begin{subfigure}[t]{0.98\textwidth}
        \centering
        \includegraphics[width=0.98\textwidth]{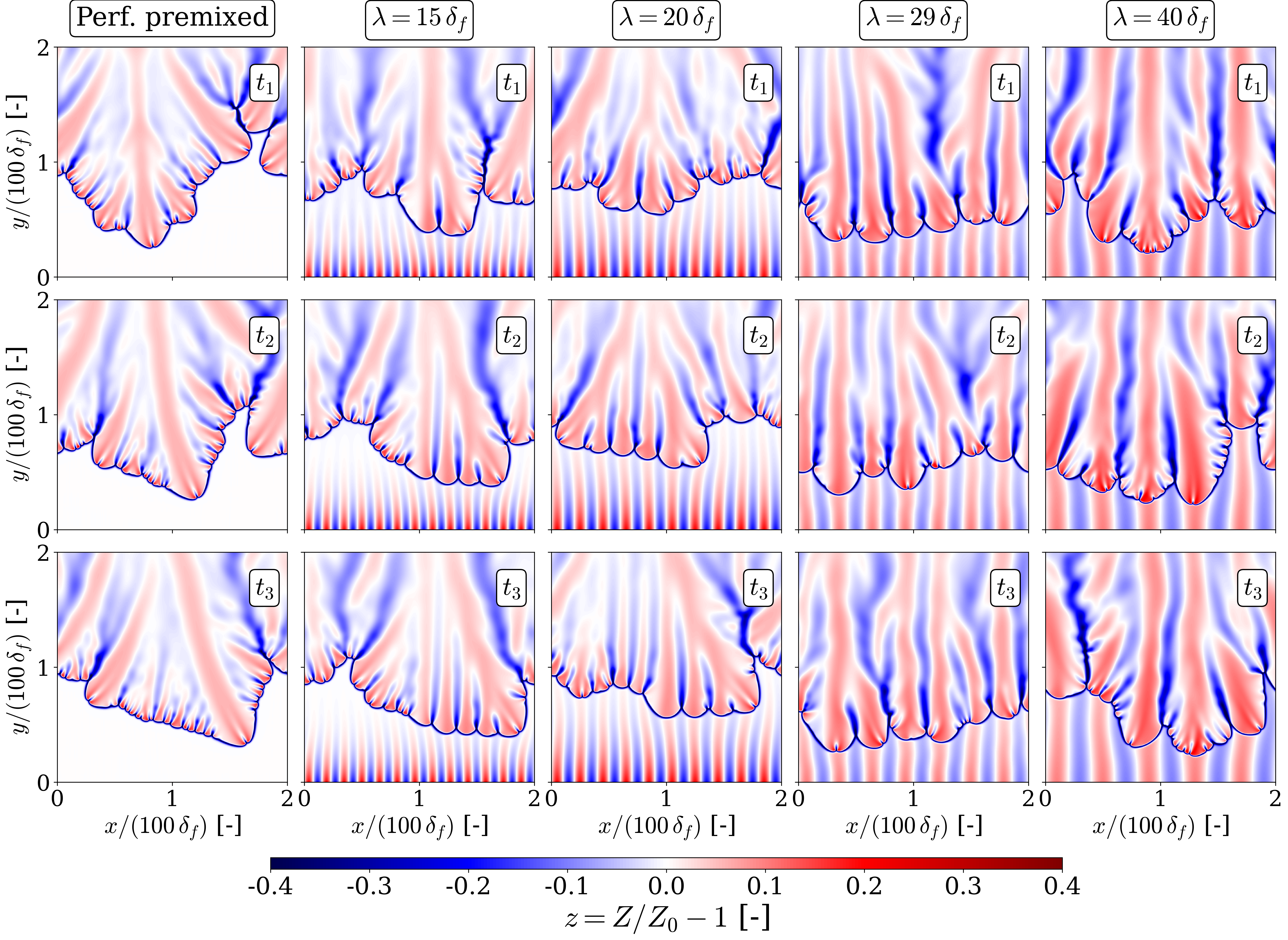}
    \end{subfigure}
    \vspace{4pt}
    \begin{subfigure}[t]{0.98\textwidth}
        \centering
        \includegraphics[width=0.98\textwidth]{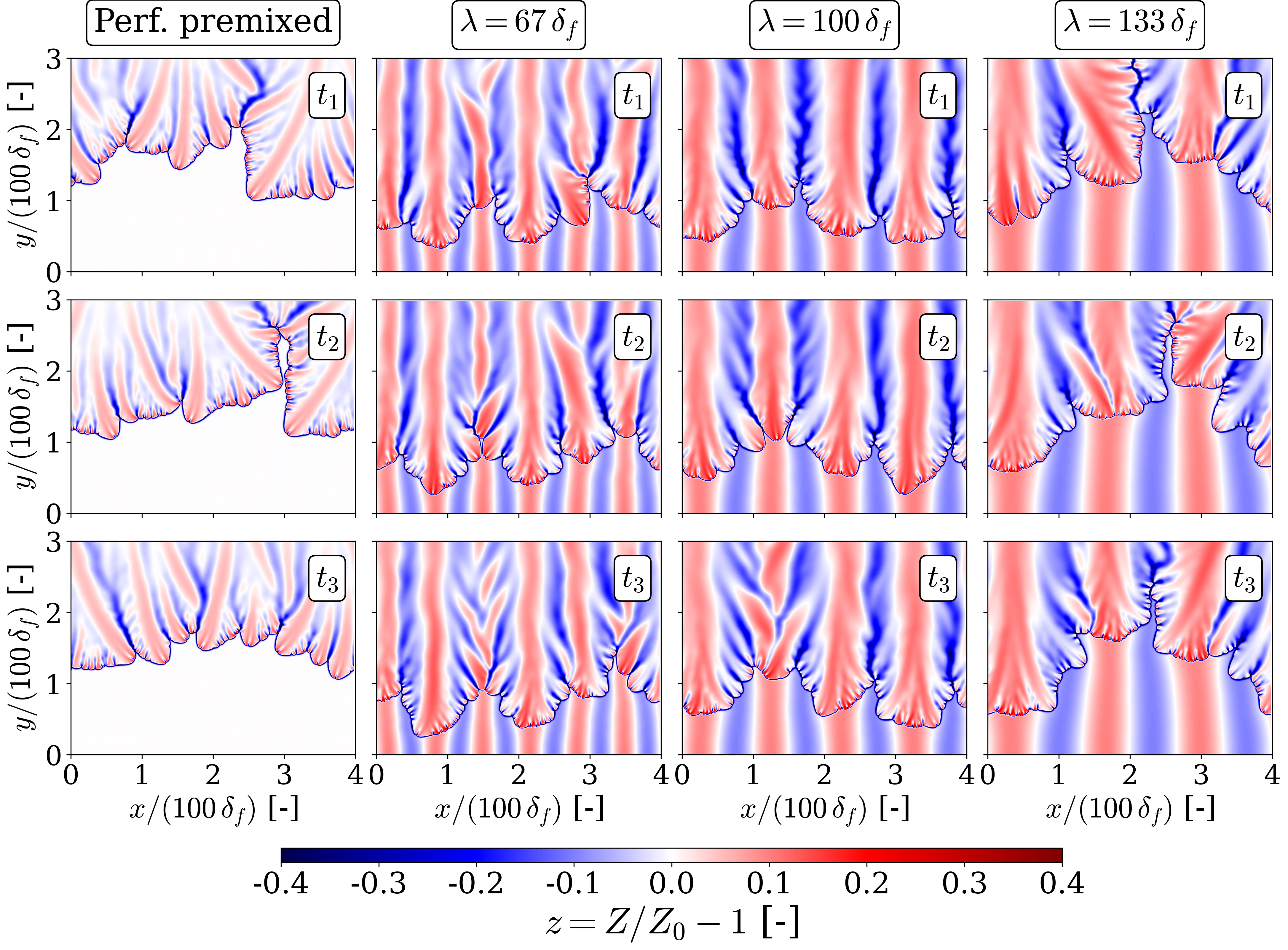}
    \end{subfigure}
    \caption{\footnotesize Instantaneous fields of the normalised Bilger mixture fraction, $z=Z/Z_0-1$, at three representative time instants ($t_1$, $t_2$, $t_3$) within the statistically stationary regime. Top: perfectly premixed reference case and $\lambda=15.4$, $20$, $28.6$, and $40\,\delta_f$. Bottom: perfectly premixed reference case and $\lambda=66.7$, $100$, and $133.3\,\delta_f$.}
    \label{fig:dynamics}
    \vspace{-10pt}
\end{figure*}

\textcolor{black}{The unity-Lewis-number configuration provides a reference to isolate the geometrical effect of the imposed stratification from thermodiffusive transport. Figure~\ref{fig:le1_stratified} shows an instantaneous field of the normalised Bilger mixture-fraction perturbation $z$ for the case $\lambda=100\,\delta_f$, together with the flame-front location identified by the $\theta=0.5$ iso-contour, where $\theta=(T-T_u)/(T_{ad}-T_u)$. The sinusoidal inlet modulation generates alternating fuel-rich and fuel-lean channels that persist up to the flame front, which responds by developing large-scale bulges in the richer regions and recessed cusps in the leaner ones. The front is therefore organised primarily at the imposed wavelength through differential local propagation speeds, yielding a hydrodynamic-like response locked to the mixture-fraction field.}

The effect of inlet fuel stratification on the flame structure in the non-unity Lewis number cases is illustrated in Fig.~\ref{fig:dynamics}, which presents instantaneous fields at three representative times for the seven analysed wavelengths together with the perfectly premixed reference case.

\textcolor{black}{In the perfectly premixed non-unity Lewis number reference, the flame displays the typical nonlinear morphology of thermodiffusively unstable lean hydrogen flames, characterised by small cellular structures that continuously form, interact, and break up, and by larger finger-like protrusions associated with cusp formation~\cite{berger2023}, as visible in Fig.~\ref{fig:dynamics}. When stratification is imposed, the response depends on the ratio between the imposed wavelength and these intrinsic scales. For the smallest wavelengths, $\lambda=15.4$--$28.6\,\delta_f$, the imposed modulation overlaps with the natural cellular range and the associated transverse gradients continuously disrupt the intrinsic thermodiffusive structures. Although the stratification is partially smoothed by transverse diffusion away from the inlet, the flame stabilises in regions where the imposed inhomogeneity remains sufficiently strong to interact with the front. As a result, small-scale wrinkles are rapidly destroyed and replaced by newly formed structures whose size is set by the imposed stratification (see Section~\ref{sec:cell_size}). This repeated breakdown and reformation prevents the persistence of coherent finger-like features, leading to a reduced surface area and a lower global propagation speed, as discussed in Section~\ref{sec:flame_speed}.}

\textcolor{black}{For larger wavelengths, $\lambda\geq40\,\delta_f$, the imposed stratification scale exceeds the characteristic size of the intrinsic thermodiffusive cells, and the flame response reflects the superposition of two mechanisms. At large scale, the flame organisation approaches that observed in the unity-Lewis-number case, with bulges forming within fuel-rich channels and cusp-like structures in lean regions. At the same time, thermodiffusive effects remain active at smaller scales, leading to additional corrugation superimposed on this large-scale structure. As a result, multiple cellular structures develop within each enriched channel and are partially guided by the imposed composition field. This combined behaviour increases the flame surface area, with the large-scale geometry set by the stratification and the fine-scale structure sustained by thermodiffusive instability. The resulting geometrical reorganisation of the flame surface is analysed in more detail in Section~\ref{sec:local}, and leads to faster global propagation relative to the perfectly premixed reference, as discussed in Section~\ref{sec:flame_speed}.}

\subsection{Flame speed analysis\label{sec:flame_speed}}\addvspace{10pt}

The global propagation speed is evaluated from the overall fuel consumption rate following~\cite{attili2021}:
\begin{equation}
s_c = -\frac{\displaystyle \int \rho\,\dot{Y}_{H_2}\,dx\,dy}{\rho_u\,Y_{H_2,u}\,L_x},
\label{eq:sc_def_mod}
\end{equation}
where $\rho$ is the local density, $\dot{Y}_{H_2}$ is the hydrogen reaction rate (in units of time inverse), $\rho_u$ and $Y_{H_2,u}$ are the unburnt density and hydrogen mass fraction, respectively, and $L_x$ is the transverse domain size. \textcolor{black}{The total flame area $\Sigma$ is computed following Vervisch et al.~\cite{vervisch1995}, with the flame front defined by the isocontour $C_{H_2}=0.89$ corresponding to the peak reaction rate of the associated one-dimensional laminar flame.} The flame speed then relates to the total flame surface as $s_c/s_L = I_0\,\Sigma/L_x$, where $s_L$ is the unstretched laminar flame speed and $I_0$ is a stretch factor accounting for local flame-stretch effects. \textcolor{black}{For each case, the flame is first allowed to evolve for 500 flame times while the mean inlet velocity is adjusted to stabilise the flame position. Statistical averages of $s_c/s_L$, $\Sigma/L_x$, and $I_0$ are then computed over at least 500 flame times after statistically stationary conditions have been reached.}

\begin{figure}[!ht]
\centering
\includegraphics[width=\columnwidth]{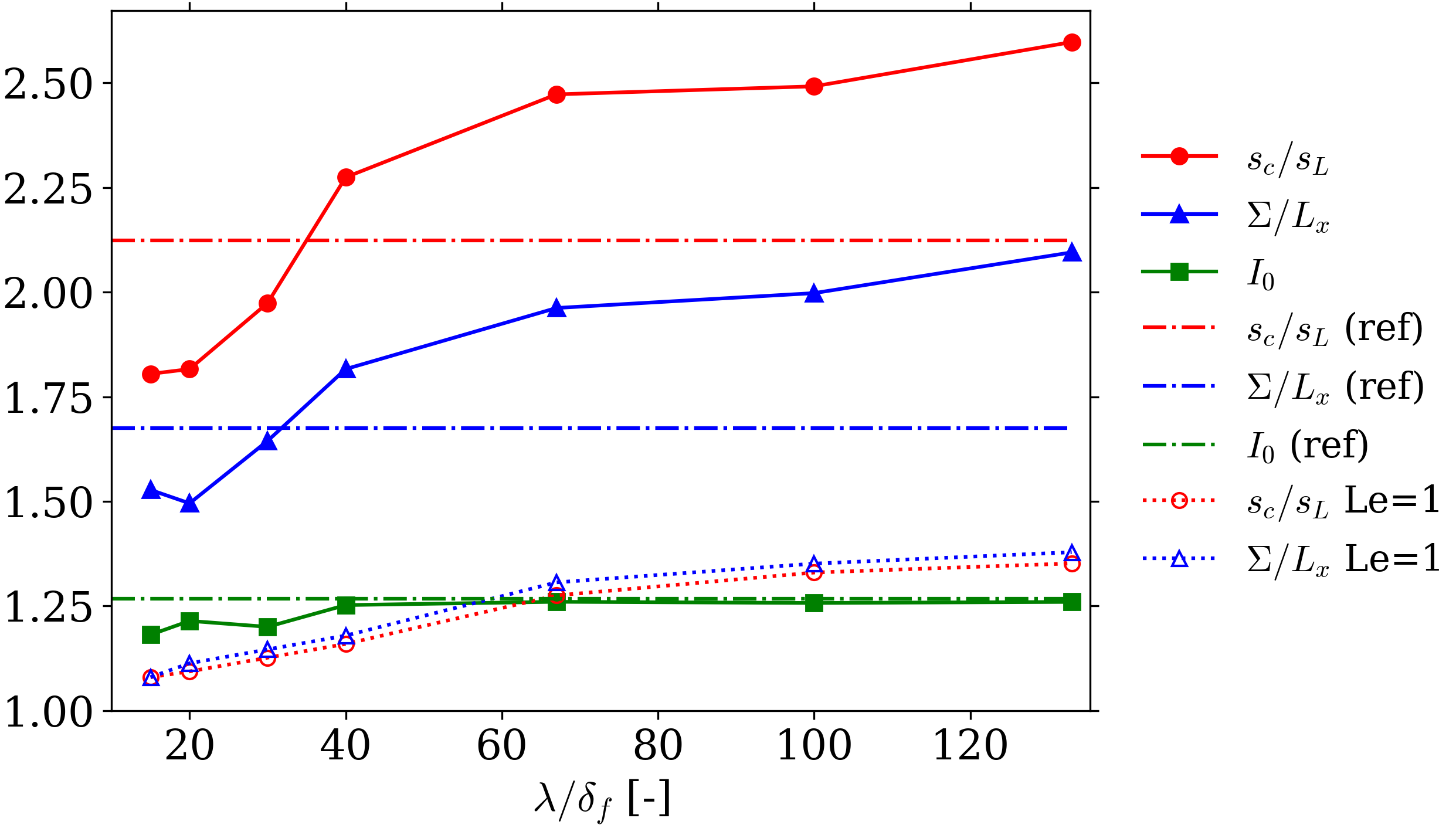}
\caption{\footnotesize Global flame properties as functions of the imposed stratification wavelength $\lambda/\delta_f$: normalised flame speed $s_c/s_L$, normalised flame surface area $\Sigma/L_x$, and stretch factor $I_0$. The horizontal dashed line indicates the perfectly premixed reference case.}
\label{fig:global}
\vspace{-10pt}
\end{figure}

\textcolor{black}{Figure~\ref{fig:global} reports these quantities as functions of the imposed wavelength for both mixture-averaged and unity-Lewis-number transport. In the unity-Lewis-number cases, the flame speed increases monotonically with $\lambda/\delta_f$, as expected from the corresponding increase in flame surface area, and remains larger than $s_L$ for all imposed wavelengths.}

\textcolor{black}{The non-unity Lewis number cases show a qualitatively similar trend for large wavelengths, $\lambda\geq40\,\delta_f$, where both $s_c/s_L$ and $\Sigma/L_x$ increase above the reference and follow the same ordering as the unity-Lewis-number cases. In this regime, $I_0$ remains approximately constant and close to the reference value, indicating that the increase in global propagation speed is primarily driven by the enlargement of the flame surface induced by the stratification, consistently with the geometrical mechanism discussed in Section~\ref{sec:dynamics}.}

\textcolor{black}{A marked difference emerges for smaller wavelengths, $\lambda\leq29\,\delta_f$. In contrast to the unity-Lewis-number behaviour, both $s_c/s_L$ and $\Sigma/L_x$ fall below the perfectly premixed reference. This reduction is accompanied by a decrease in $I_0$, indicating that the imposed small-scale stratification not only reduces the flame surface area but also weakens the local burning enhancement associated with stretch and differential diffusion. The combined effect leads to propagation speeds lower than the reference case, highlighting the disruptive action of transverse composition gradients on both large thermodiffusive structures and local flame dynamics.}

\subsection{Flame-cell size distribution\label{sec:cell_size}}\addvspace{10pt}

\textcolor{black}{To quantify the redistribution of flame surface among cellular structures of different size, flame cells are extracted along the isocontour of the hydrogen-based progress variable corresponding to the peak reaction rate in the associated one-dimensional laminar flame, $C_{H_2}=0.89$. Along this contour, individual cells are identified as segments bounded by neighbouring cusps. For each cell, the chord length $l$ is defined as the straight-line distance between the two cusp points delimiting the cell, while the corresponding arc length $s$ is the length of the contour between the same two points.}

\textcolor{black}{Each cell is then classified according to its chord length normalised by the laminar flame thickness, $l/\delta_f$. For each instantaneous flame contour, the distribution is constructed using an arc-length weighting, so that the fraction of flame length associated with bin $i$ is given by $f_i(t)=\sum_{j \in i} s_j(t)\big/\sum_j s_j(t)$, where $s_j(t)$ is the arc length of cell $j$, and the numerator includes only cells whose chord length falls within bin $i$. Thus, $f_i(t)$ represents the fraction of the instantaneous flame length occupied by cells in a given chord-length range. The mean length-fraction distribution $\overline{f}_i$ is obtained by averaging over $N_t$ statistically independent snapshots in the stationary regime. To quantify deviations from the perfectly premixed reference, the relative change in each bin is defined as $\Delta \overline{f}_{i} = (\overline{f}_{i,\mathrm{case}} - \overline{f}_{i,\mathrm{ref}})/\overline{f}_{i,\mathrm{ref}}$.}

\begin{figure}[!ht]
    \centering
    \begin{subfigure}[t]{\columnwidth}
        \centering
        \includegraphics[width=\columnwidth]{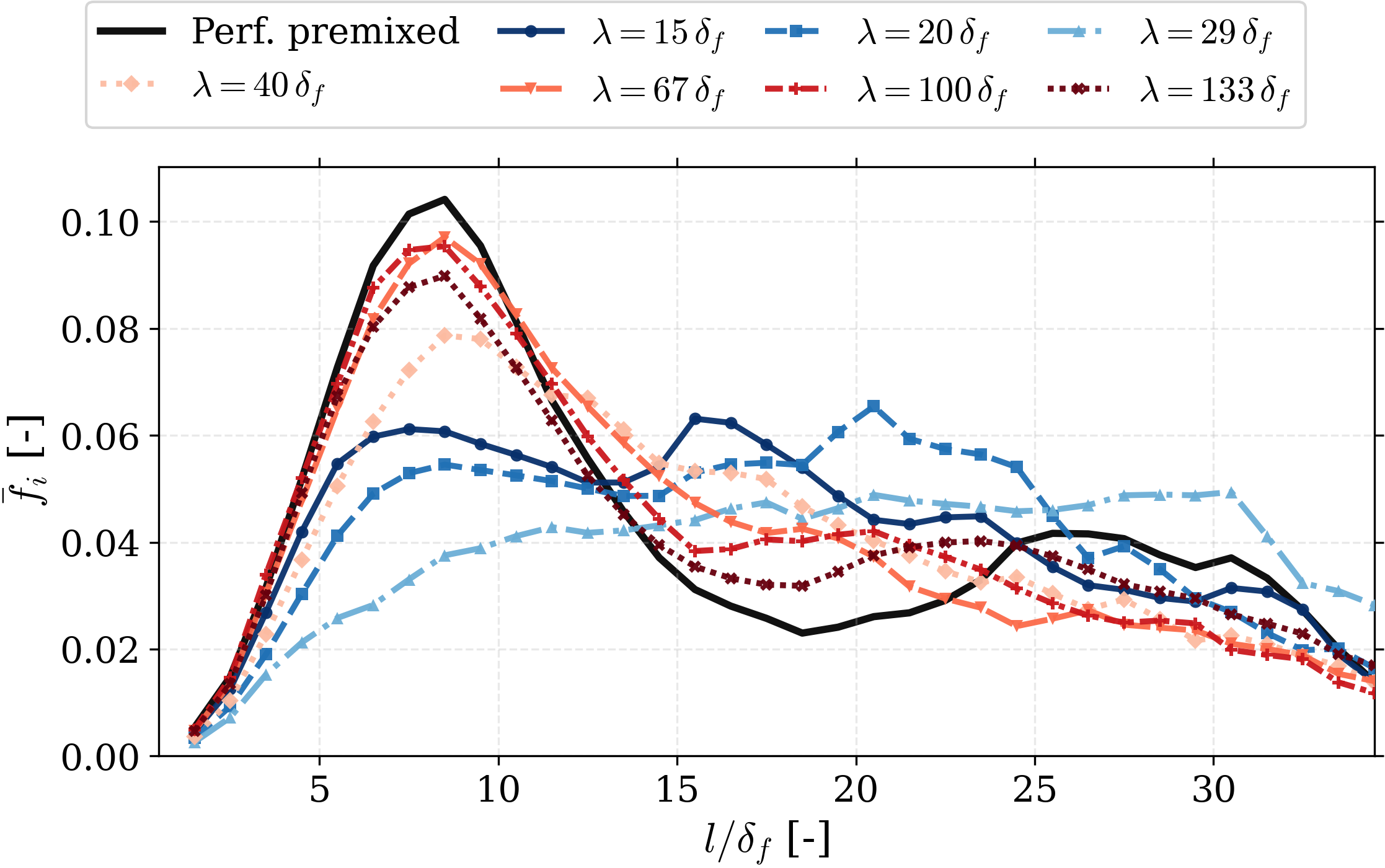}
    \end{subfigure}
    \vspace{4pt}
    \begin{subfigure}[t]{\columnwidth}
        \centering
        \includegraphics[width=\columnwidth]{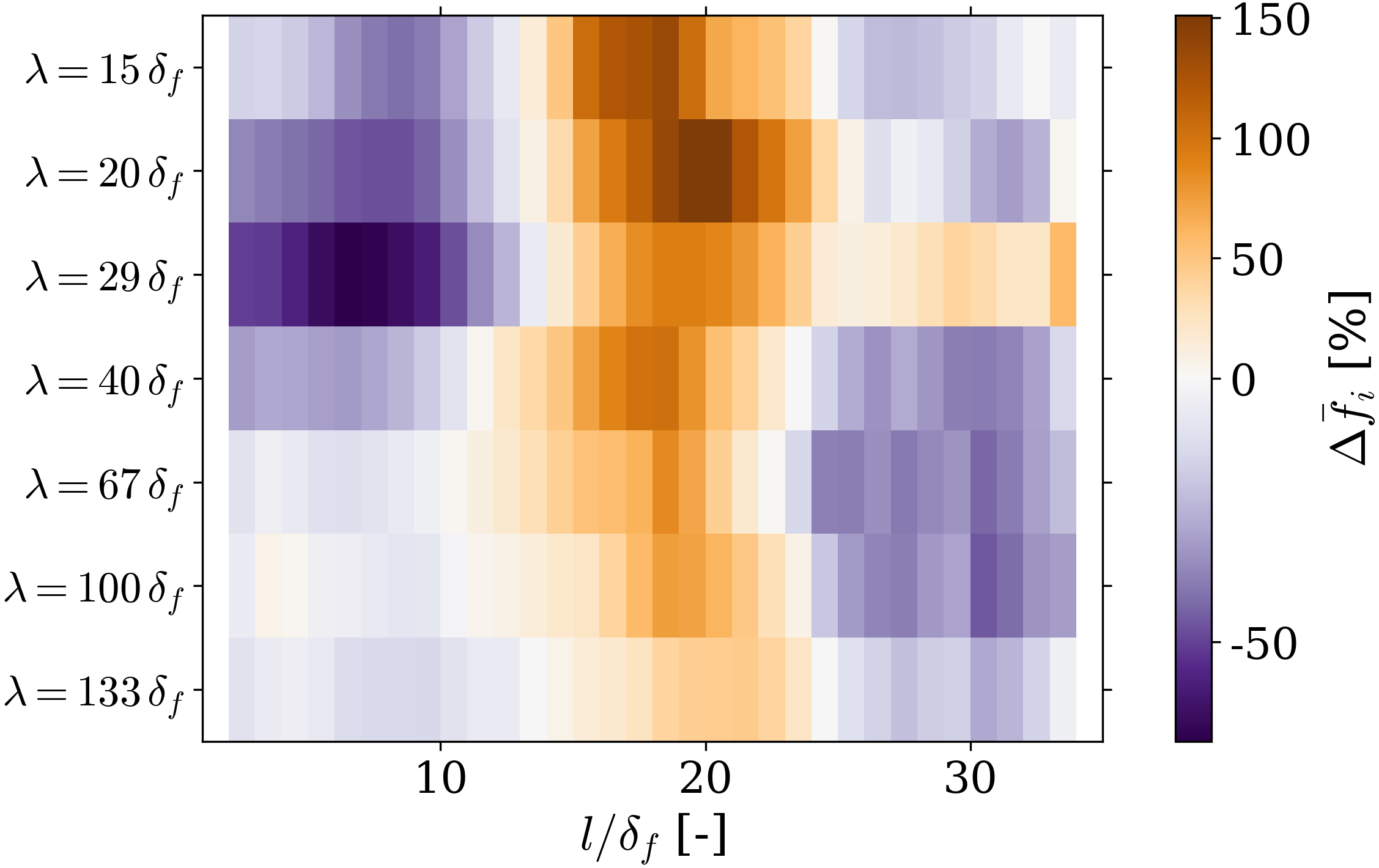}
    \end{subfigure}
    \caption{\footnotesize Flame-cell size statistics. Top: arc-length-weighted distribution of cell chord length, $l/\delta_f$. Bottom: relative variation with respect to the perfectly premixed reference, $\Delta \overline{f}_{i}$.}
    \label{fig:cell_distribution}
    \vspace{-10pt}
\end{figure}

\textcolor{black}{To characterise how inlet stratification modifies the cellular structure of the flame, Fig.~\ref{fig:cell_distribution} reports the distribution of flame length across cell chord lengths, $l/\delta_f$, together with the corresponding variation relative to the perfectly premixed reference. The upper panel shows the arc-length-weighted distribution of cell chord length, while the lower panel reports the corresponding relative variation with respect to the perfectly premixed reference, $\Delta \overline{f}_{i}$, highlighting enriched and depleted chord-length ranges.}

\textcolor{black}{The perfectly premixed reference provides the baseline for the intrinsic nonlinear cell dynamics. Its bimodal distribution, with peaks at $l/\delta_f \simeq 7$--$10$ and $l/\delta_f \simeq 25$--$30$, reflects two characteristic contributions of fully developed thermodiffusively unstable lean hydrogen flames. The first peak is associated with the intrinsic small cellular wrinkles, while the second corresponds to larger cusp-bounded cells produced by nonlinear interaction, widening, and merging of neighbouring structures~\cite{berger2019,berger2023}.}

\textcolor{black}{In contrast, the present stratified cases introduce an external length scale through the imposed composition field, allowing the effect of stratification to be interpreted as a modification of the natural cell-selection process. This is analogous, in a different setting, to the scale dependence observed in homogeneous unstable flames, where the flame structure and propagation depend on the number of unstable wavelengths supported by the domain~\cite{creta2020,lapenna2021proci,lapenna2021CTM}.}

\textcolor{black}{For small stratification wavelengths, $\lambda=15.4$--$28.6\,\delta_f$, the imposed scale lies within the range over which the intrinsic cellular dynamics develop. In this regime, the reference distribution is strongly altered: flame length is depleted around the intrinsic small-cell peak and redistributed towards intermediate chord lengths, with the distribution developing a peak close to the imposed wavelength, as shown in Fig.~\ref{fig:cell_distribution}. The relative-change map shows that this redistribution is not a simple broadening of the reference distribution, but a shift of flame length away from the self-selected cellular scales towards structures controlled by the imposed modulation. This indicates that transverse composition gradients interfere directly with the natural formation, widening, and splitting of thermodiffusive cells. Existing cells are repeatedly disrupted, and new cusp-bounded structures are preferentially formed at scales dictated by the stratification rather than by the unforced nonlinear dynamics. This explains the reduction of $I_0$ observed in Section~\ref{sec:flame_speed}.}

\textcolor{black}{For larger wavelengths, $\lambda\geq40\,\delta_f$, the imposed scale is sufficiently separated from the intrinsic small-cell range. The small-cell peak is largely preserved, indicating that thermodiffusive cellular dynamics remain active. Deviations from the reference are weaker overall and primarily concentrated at larger chord lengths, where the imposed rich and lean channels reorganise the flame front into bulges and cusps. Thus, in this regime, stratification does not replace the intrinsic cell-selection process; it superimposes a large-scale geometrical modulation on a flame front that still retains its thermodiffusive cellular structure, in agreement with the nearly unchanged $I_0$ reported in Section~\ref{sec:flame_speed}.}

\textcolor{black}{Overall, Fig.~\ref{fig:cell_distribution} shows that the imposed stratification changes the role of the nonlinear cell dynamics identified in homogeneous flames. At small wavelengths, the imposed stratification directly interferes with the intrinsic cell-selection process, promoting the emergence of structures at the imposed scale. At larger wavelengths, it leaves the intrinsic cellular range largely intact and induces weaker modifications to the shape of the distribution.}

\subsection{Local analysis of flame--stratification interaction\label{sec:local}}\addvspace{10pt}

The global metrics discussed in Section~\ref{sec:flame_speed} show that inlet fuel stratification with $\lambda\geq40\,\delta_f$ primarily modifies flame propagation through changes in the total flame surface area. While this global analysis identifies the dominant role of flame geometry, it does not clarify the local mechanisms by which the stratified mixture reorganises the flame surface. To address this question, the flame dynamics are analysed locally relative to the stratified mixture field described by the normalised mixture-fraction perturbation $z$ introduced in Section~\ref{sec:dynamics}.

A difficulty arises because the mixture fraction does not remain constant across the flame structure. \textcolor{black}{In the present stratified flames, the mixture fraction within the reaction zone reflects both the imposed inlet inhomogeneity and flame-induced differential-diffusion effects. Conditioning directly on $Z$ would therefore mix the upstream stratification with local modifications generated inside the flame. For this reason, the passive tracer is used to label the transported stratification channels independently of chemical source terms and differential diffusion.}

To overcome this limitation, a passive scalar tracer $Z_{tr}$ is introduced to track the inlet stratification independently of combustion chemistry. The tracer obeys the same advection--diffusion equation as other scalars but does not participate in chemical reactions and is transported with $Le=0.3$. This allows the large-scale stratification pattern to be followed as it is convected through the domain without being affected by chemical source terms or differential diffusion.

At the inlet, the tracer is prescribed with the same sinusoidal modulation used for the fuel stratification,
\begin{equation}
Z_{tr}(x) = Z_{tr,0}\,[\,1 + A\sin(2\pi x/\lambda)\,],
\end{equation}
where $Z_{tr,0}=1$ and $A$ is set equal to the modulation amplitude of the corresponding case. For consistency with the definition of $z$, the tracer field is expressed as $z_{tr} = Z_{tr}/Z_{tr,0} - 1$. The tracer therefore acts as a marker of the large-scale stratification channels while remaining unaffected by reactions and differential diffusion.

\begin{figure}[!ht]
\centering
\includegraphics[width=0.8\columnwidth]{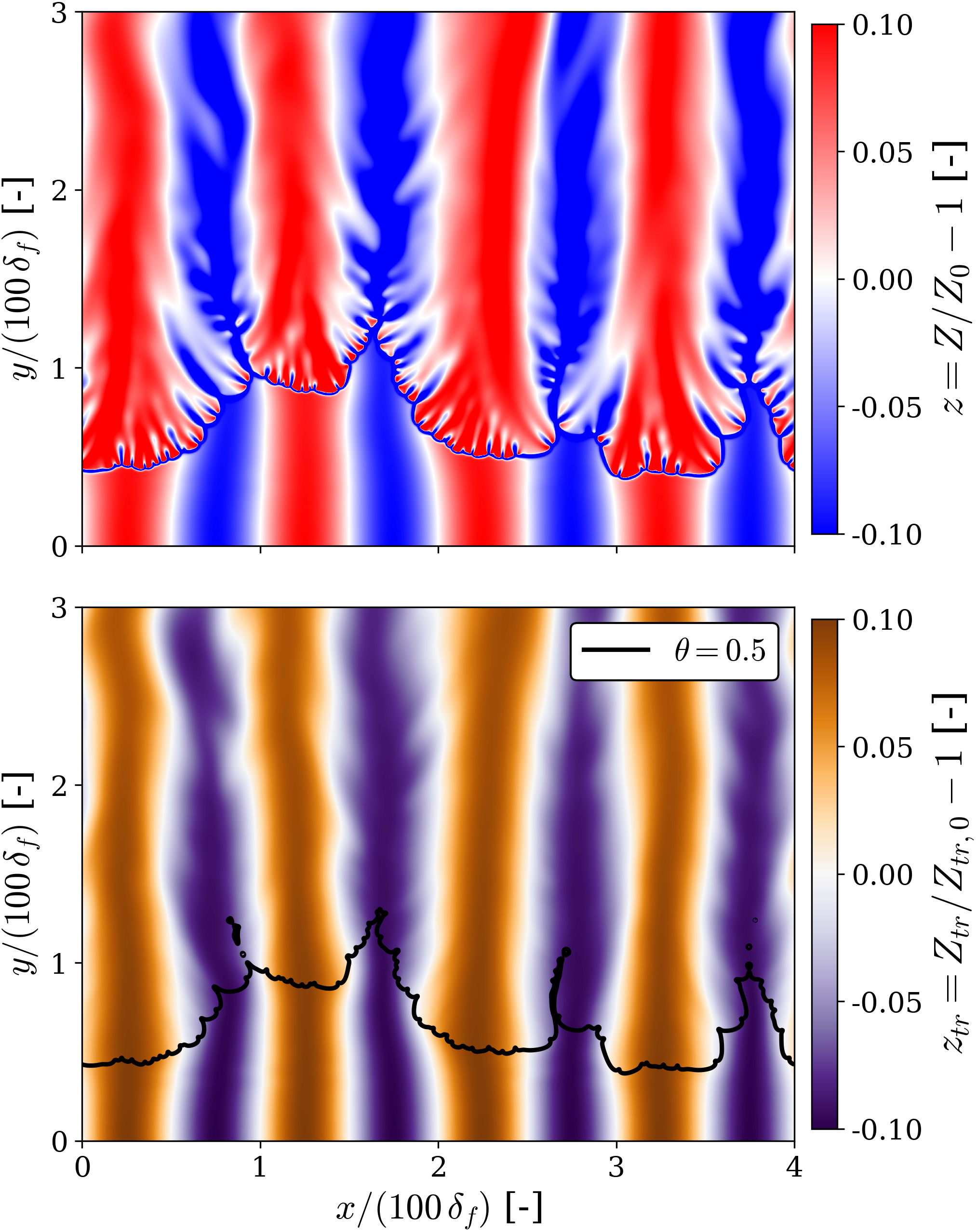}
\caption{\footnotesize Instantaneous fields of the mixture-fraction perturbation $z$ (top) and passive tracer $z_{tr}$ (bottom). The black contour denotes the iso-line $\theta=0.5$, indicating the flame location.}
\label{fig:z_vs_ztr}
\vspace{-10pt}
\end{figure}

Figure~\ref{fig:z_vs_ztr} compares instantaneous fields of the mixture-fraction perturbation $z$ and of the passive tracer $z_{tr}$. The imposed stratification pattern is strongly distorted in the $z$ field as the flow crosses the reaction zone due to differential diffusion effects. In contrast, the tracer preserves the large-scale fuel-rich and fuel-lean channels throughout the domain, allowing the convective transport of the stratification pattern to be clearly identified and providing a clearer representation of the mixture field experienced by the flame.

\begin{figure*}[!t]
\centering
\includegraphics[width=0.85\textwidth]{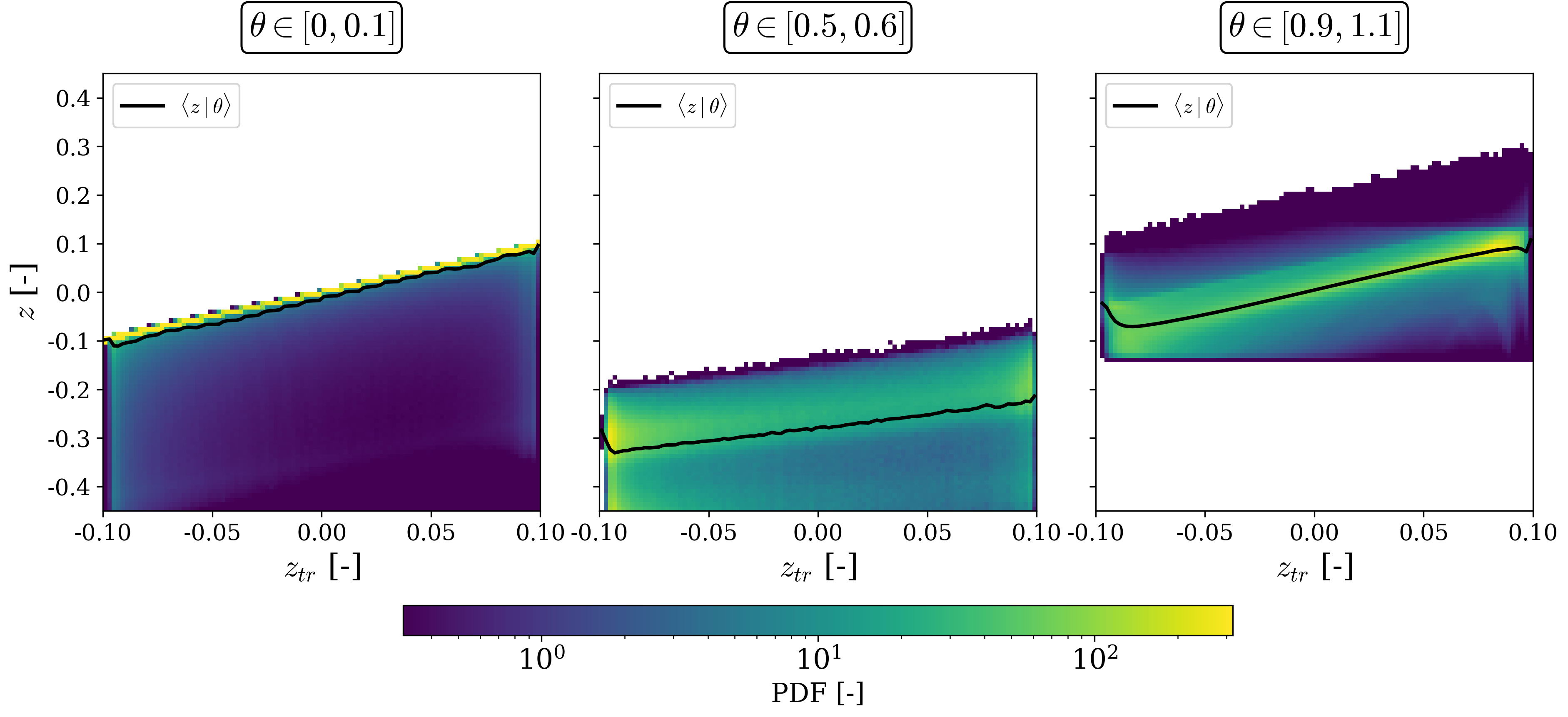}
\caption{\footnotesize Joint probability density functions of $z$ and $z_{tr}$ conditioned on three intervals of the normalised temperature $\theta$, corresponding to unburnt gases (left), the reaction zone (middle), and burnt gases (right). The black line denotes the conditional mean $\langle z \mid z_{tr}\rangle$. The case shown corresponds to $\lambda=100\,\delta_f$.}
\label{fig:z_jpdf}
\vspace{-10pt}
\end{figure*}

\textcolor{black}{This behaviour is quantified in Fig.~\ref{fig:z_jpdf}, which reports JPDFs of $z$ and $z_{tr}$ for $\lambda=100\,\delta_f$, conditioned on three intervals of the normalised temperature: the unburnt mixture ($\theta\in[0,0.1]$), the reaction zone ($\theta\in[0.5,0.6]$), and the burnt gases ($\theta\in[0.9,1.1]$). In the unburnt mixture, the distributions remain close to the diagonal, indicating that $z_{tr}$ follows the convected stratification pattern upstream of the flame. Within the reaction zone, the correlation weakens because differential diffusion modifies the local mixture fraction. In the burnt gases, the distribution collapses again close to the diagonal, showing that the mixture fraction largely recovers the large-scale stratification pattern after crossing the flame. The tracer is therefore used in the following as the conditioning coordinate to identify the rich and lean channels interacting with the flame.}

\begin{figure}[!ht]
\centering
\includegraphics[width=\columnwidth]{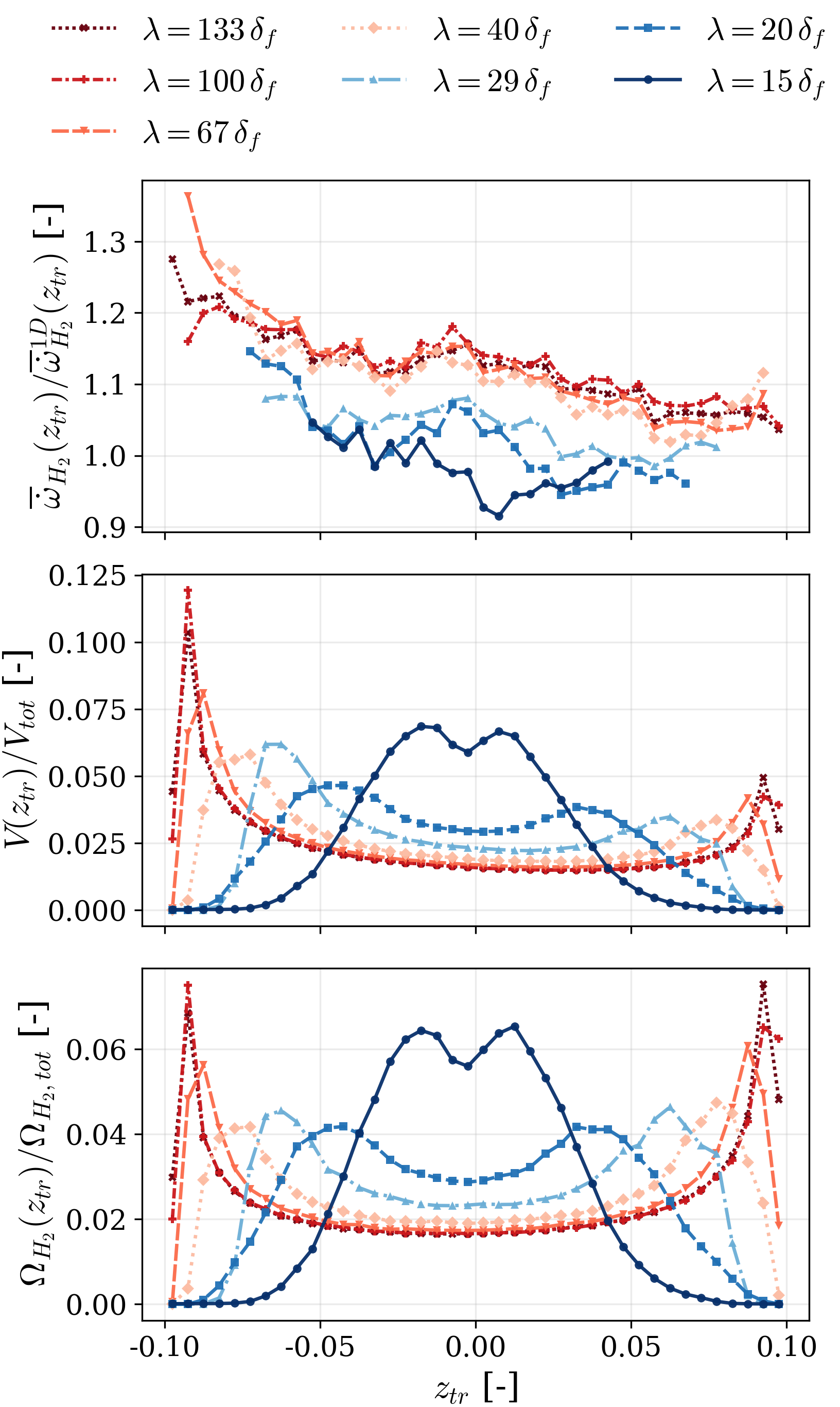}
\caption{\footnotesize Distribution of burning activity across the stratified mixture as a function of the tracer coordinate $z_{tr}$ for all investigated wavelengths. Top: mean reaction rate $\bar{\dot{\omega}}_{H_2}(z_{tr})$ \textcolor{black}{normalised by the corresponding mean one-dimensional flamelet value}. Middle: fraction of reactive volume $V(z_{tr})/V_{tot}$. Bottom: fraction of total hydrogen consumption $\Omega(z_{tr})/\Omega_{tot}$.}
\vspace{-10pt}
\label{fig:global_omega_condmean}
\end{figure}

Figure~\ref{fig:global_omega_condmean} reports the distribution of burning activity across the imposed stratification using $z_{tr}$ as conditioning coordinate. The notation $(z_{tr})$ denotes quantities evaluated within bins of the tracer coordinate. \textcolor{black}{For each $z_{tr}$ bin, the reactive volume is defined as the set of grid cells where the local hydrogen consumption rate exceeds 1\% of the maximum value obtained from the corresponding one-dimensional laminar flame at the local mixture composition associated with that bin. This local flamelet-based definition accounts for the variation of laminar flame thickness and reaction-rate magnitude across the stratification.} Accordingly, $V(z_{tr})$ is the reactive volume associated with a given tracer bin, $V_{tot}$ is the total reactive volume, $\Omega(z_{tr})$ is the hydrogen consumption rate integrated over the reactive cells in that bin, and $\Omega_{tot}$ is the corresponding integral over the full reactive volume. The top panel shows the mean hydrogen consumption rate $\bar{\dot{\omega}}_{H_2}(z_{tr})$, \textcolor{black}{normalised by the mean one-dimensional flamelet value $\bar{\dot{\omega}}_{H_2}^{1D}(z_{tr})$, with both means evaluated over the corresponding reactive volume. Only $z_{tr}$ bins with sufficient statistical support, here defined by $V(z_{tr})/V_{tot}>0.01$, are shown in this panel.} The middle panel shows the fraction of reactive volume $V(z_{tr})/V_{tot}$. The bottom panel shows the fraction of total hydrogen consumption $\Omega(z_{tr})/\Omega_{tot}$.

\begin{figure*}[!t]
\centering
\includegraphics[width=0.85\textwidth]{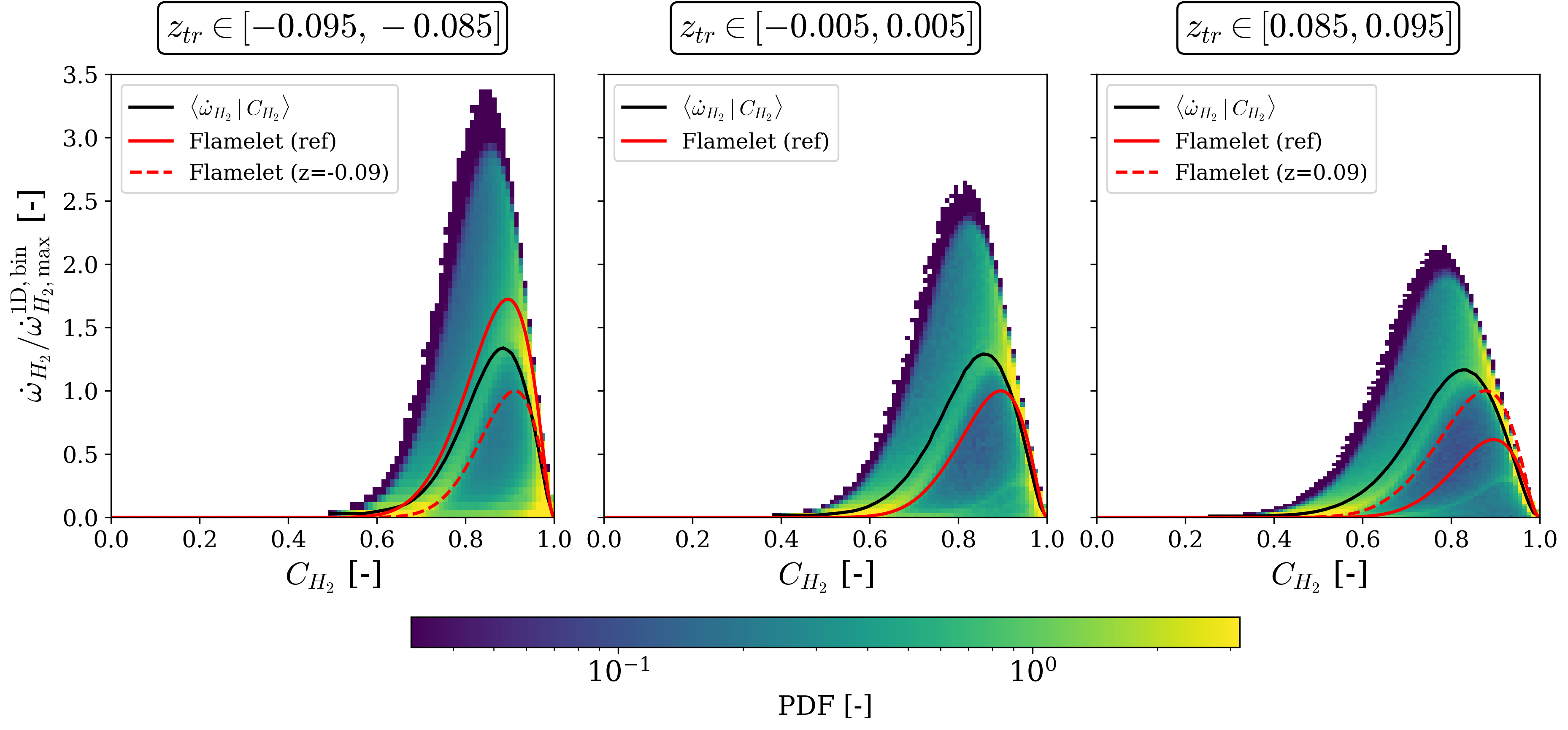}
\caption{\footnotesize JPDFs of the hydrogen consumption rate $\dot{\omega}_{H_2}$ versus the hydrogen-based progress variable $C_{H_2}$, conditioned on lean, nominal, and rich tracer bins for the case $\lambda=100\,\delta_f$. The reaction rate is normalised by the maximum value of the corresponding one-dimensional flamelet for each bin. The black line denotes the conditional mean $\langle \dot{\omega}_{H_2}\mid C_{H_2}\rangle$. Red curves show the corresponding laminar flamelet profiles.}
\label{fig:omega_jpdf}
\vspace{-10pt}
\end{figure*}

\textcolor{black}{The reduced values of $V(z_{tr})/V_{tot}$ in the outermost tracer bins indicate that the flame samples the most extreme rich and lean stratification levels only over a limited reactive volume. This is a consequence of the diffusive smoothing of $z_{tr}$ during convection from the inlet to the flame, which reduces the range of tracer values reaching the reaction zone. The effect is stronger for smaller wavelengths, where transverse diffusion is more effective.}

At small wavelengths, the local mean reaction-rate ratio is reduced across all $z_{tr}$, indicating an overall weakening of the flame consistent with the global trends of Section~\ref{sec:flame_speed}. For all wavelengths, the ratio decreases weakly from lean to rich tracer bins, indicating a small residual preferential-diffusion enhancement on the lean side. The reactive volume is preferentially distributed towards lean regions: for large wavelengths, rich channels form forward-propagating bulges, while lean channels develop recessed cusps that concentrate a significant fraction of the flame surface. As a result, \textcolor{black}{the integrated consumption reflects the balance between two effects: richer regions have larger absolute burning rates due to their local mixture composition, whereas leaner regions contribute through a larger reactive volume. This balance leads} to a nearly symmetric distribution of total consumption across $z_{tr}$. This symmetry allows the faster rich-channel structures and the slower lean-channel cusps to remain dynamically coupled, preventing fragmentation of the flame front.

\textcolor{black}{To assess the role of preferential diffusion in the local burning behaviour, Fig.~\ref{fig:omega_jpdf} reports JPDFs of the hydrogen consumption rate $\dot{\omega}_{H_2}$ as a function of the hydrogen-based progress variable $C_{H_2}$, conditioned on three tracer bins representative of lean ($z_{tr}\in[-0.095,-0.085]$), nominal ($z_{tr}\in[-0.005,0.005]$), and rich ($z_{tr}\in[0.085,0.095]$) regions. The case shown corresponds to $\lambda=100\,\delta_f$. The reaction rate is normalised by the maximum value of the corresponding one-dimensional flamelet for each tracer bin, so that the leading-order dependence on the local equivalence ratio is removed.}

Under this normalisation, the conditional means \textcolor{black}{remain close across the three tracer bins, although they do not collapse exactly}. \textcolor{black}{The variations are limited, remaining below approximately $10\%$ over the main reaction-rate region, consistently with the relatively weak tracer dependence observed in Fig.~\ref{fig:global_omega_condmean}.} This indicates that the mean local burning response is primarily governed by the local mixture composition. The JPDFs nevertheless exhibit a slightly broader spread in the lean bin, reflecting residual variability associated with preferential diffusion. However, this dispersion \textcolor{black}{only produces a secondary modulation and} does not significantly alter the mean behaviour.

These results indicate that, for the moderate stratification amplitudes considered here, differential-diffusion effects introduce a secondary modulation of the local burning dynamics, while the dominant dependence remains controlled by the imposed variation in mixture composition. Combined with the tracer-conditioned analysis of Fig.~\ref{fig:global_omega_condmean}, this supports the interpretation that the increase in global flame speed at large wavelengths originates primarily from the geometrical redistribution of the flame surface, rather than from a systematic enhancement of local burning rates.

\section{Conclusions\label{sec:conclusions}}\addvspace{10pt}

This work investigated the effect of inlet fuel stratification on lean premixed hydrogen flames subjected to thermodiffusive instabilities using DNS with detailed chemistry and transport. Controlled sinusoidal perturbations of the fuel mass fraction were imposed at the inlet to systematically vary the characteristic wavelength of the stratification and analyse its interaction with the intrinsic thermodiffusive instability of lean hydrogen flames. \textcolor{black}{Moderate stratification amplitudes are considered, leading to local equivalence-ratio variations of approximately $\phi\simeq0.45$--$0.55$ in the flame region.}

The results reveal a strongly scale-dependent flame response, governed by the competition between the imposed stratification wavelength and the natural length scales selected by thermodiffusive instability. \textcolor{black}{For small wavelengths, $\lambda=15.4$--$28.6\,\delta_f$, the imposed mixture inhomogeneity acts at scales comparable to the intrinsic cellular dynamics. In this regime, transverse composition gradients disrupt the formation and persistence of thermodiffusive cells, redistribute flame length away from the self-selected cellular scales, and reduce both the flame surface area and the global propagation speed. The associated decrease of the stretch factor $I_0$ further indicates that small-scale stratification weakens not only the flame geometry but also the local burning enhancement produced by preferential diffusion.}

For larger wavelengths, $\lambda\geq40\,\delta_f$, the flame response changes qualitatively. The imposed stratification organises the flame into alternating fuel-rich and fuel-lean channels, similarly to the corresponding unity-Lewis-number cases, while thermodiffusive cells remain active at smaller scales. This results in forward-propagating bulges in rich regions and recessed cusp-like structures in lean regions. The resulting geometrical reorganisation increases the total flame surface area and enhances the global propagation speed, with $I_0$ remaining close to the perfectly premixed reference. \textcolor{black}{Local analysis indicates that the absolute burning rate is primarily controlled by the local mixture composition, while differential diffusion introduces only a secondary modulation. Tracer-conditioned statistics show that leaner regions occupy a larger fraction of the reactive volume through recessed cusp-like structures, whereas richer regions retain higher absolute burning intensities associated with their local composition. These competing effects lead to a near compensation in the integrated consumption.}

Overall, these findings identify the stratification length scale as a key parameter governing the interaction between mixture inhomogeneity and thermodiffusive instability in lean hydrogen flames. \textcolor{black}{The results show that fuel stratification can either suppress or reinforce large-scale flame structures depending on its characteristic scale, providing a framework to interpret how imposed inhomogeneities interact with intrinsic instability mechanisms.} \textcolor{black}{These insights are relevant for the modelling and control of hydrogen combustion systems where incomplete premixing or upstream inhomogeneities are present.} \textcolor{black}{Larger stratification amplitudes may strengthen the coupling between local composition, density variations, preferential diffusion, and flame geometry; therefore, the present results should be regarded as isolating the scale-selection mechanism at moderate amplitude, while the extension to stronger mixture variations requires dedicated investigation.}

\acknowledgement{CRediT authorship contribution statement} \addvspace{10pt}

{\bf F.\ Fruzza}: Conceptualization; Methodology; Investigation; Writing – Original draft. {\bf S.\ Al Kassar}: Methodology; Software; Writing – Review \& editing. {\bf R.\ Lamioni}: Writing – Review \& editing. {\bf C.\ Galletti}: Supervision; Writing – Review \& editing. {\bf A.\ Attili}: Conceptualization; Methodology; Supervision; Writing – Review \& editing.

\acknowledgement{Declaration of competing interest} \addvspace{10pt}

The authors declare that they have no known competing financial interests or personal relationships that could have appeared to influence the work reported in this paper.

\acknowledgement{Acknowledgments} \addvspace{10pt}

This work was supported by a Short-Term Scientific Mission Grant from the COST Action CA22151 CYPHER. This work used the ARCHER2 UK National Supercomputing Service (https://www.archer2.ac.uk). We acknowledge the CINECA award under the ISCRA initiative for the availability of high-performance computing resources and support.

\footnotesize
\baselineskip 9pt

\clearpage
\thispagestyle{empty}
\bibliographystyle{proci}
\bibliography{Collection}


\newpage

\small
\baselineskip 10pt


\end{document}